\documentstyle[psfig]{mn}

\newif\ifAMStwofonts

\ifoldfss
  \newcommand{\rmn}[1] {{\rm #1}}

  \ifCUPmtlplainloaded \else
    \NewTextAlphabet{textbfit} {cmbxti10} {}
    \NewTextAlphabet{textbfss} {cmssbx10} {}
    \NewMathAlphabet{mathbfit} {cmbxti10} {} 
    \NewMathAlphabet{mathbfss} {cmssbx10} {} 
  \fi
  \ifAMStwofonts
    \ifCUPmtlplainloaded \else
      \NewSymbolFont{upmath} {eurm10}
      \NewSymbolFont{AMSa} {msam10}
      \NewMathSymbol{\upi}     {0}{upmath}{19}
      \NewMathSymbol{\umu}     {0}{upmath}{16}
      \NewMathSymbol{\upartial}{0}{upmath}{40}
      \NewMathSymbol{\leqslant}{3}{AMSa}{36}
      \NewMathSymbol{\geqslant}{3}{AMSa}{3E}

      \let\geq=\geqslant 
    \fi
  \fi
\fi 

\ifnfssone
  \newmathalphabet{\mathit}
  \addtoversion{normal}{\mathit}{cmr}{m}{it}
  \addtoversion{bold}{\mathit}{cmr}{bx}{it}
  \newcommand{\rmn}[1] {\mathrm{#1}}

  \newmathalphabet{\mathbfit} 
  \addtoversion{normal}{\mathbfit}{cmr}{bx}{it}
  \addtoversion{bold}{\mathbfit}{cmr}{bx}{it}
  \newmathalphabet{\mathbfss} 
  \addtoversion{normal}{\mathbfss}{cmss}{bx}{n}
  \addtoversion{bold}{\mathbfss}{cmss}{bx}{n}
  \ifAMStwofonts
    \ifCUPmtlplainloaded \else
      \UseAMStwoboldmath
      \makeatletter
      \new@mathgroup\upmath@group
      \define@mathgroup\mv@normal\upmath@group{eur}{m}{n}
      \define@mathgroup\mv@bold\upmath@group{eur}{b}{n}
      \edef\UPM{\hexnumber\upmath@group}
      \new@mathgroup\amsa@group
      \define@mathgroup\mv@normal\amsa@group{msa}{m}{n}
      \define@mathgroup\mv@bold\amsa@group{msa}{m}{n}
      \edef\AMSa{\hexnumber\amsa@group}
      \makeatother
      \mathchardef\upi="0\UPM19
      \mathchardef\umu="0\UPM16
      \mathchardef\upartial="0\UPM40
      \mathchardef\leqslant="3\AMSa36
      \mathchardef\geqslant="3\AMSa3E

      \let\geq=\geqslant 
    \fi
  \fi
\fi 

\ifnfsstwo
  \newcommand{\rmn}[1] {\mathrm{#1}}

  \DeclareMathAlphabet{\mathbfit}{OT1}{cmr}{bx}{it}
  \SetMathAlphabet\mathbfit{bold}{OT1}{cmr}{bx}{it}
  \DeclareMathAlphabet{\mathbfss}{OT1}{cmss}{bx}{n}
  \SetMathAlphabet\mathbfss{bold}{OT1}{cmss}{bx}{n}
  \ifAMStwofonts
    \ifCUPmtlplainloaded \else
      \DeclareSymbolFont{UPM}{U}{eur}{m}{n}
      \SetSymbolFont{UPM}{bold}{U}{eur}{b}{n}
      \DeclareSymbolFont{AMSa}{U}{msa}{m}{n}
      \DeclareMathSymbol{\upi}{0}{UPM}{"19}
      \DeclareMathSymbol{\umu}{0}{UPM}{"16}
      \DeclareMathSymbol{\upartial}{0}{UPM}{"40}
      \DeclareMathSymbol{\leqslant}{3}{AMSa}{"36}
      \DeclareMathSymbol{\geqslant}{3}{AMSa}{"3E}

      \let\geq=\geqslant 
    \fi
  \fi
\fi 

\ifCUPmtlplainloaded \else
  \ifAMStwofonts \else 
    \def\upi{\pi}
    \def\umu{\mu}
    \def\upartial{\partial}
  \fi
\fi

\title{The Fragmentation of Pre-enriched Primordial Objects}
\author[V. Bromm {\rm et al.}]
       {V. Bromm$^{1,3}$, A. Ferrara$^2$, P. S. Coppi$^3$ and R. B. Larson$^3$ \\
      $^1$Institute of Astronomy, Madingley Road, Cambridge CB3 0HA, England\\
$^2$Osservatorio Astrofisico di Arcetri, Largo E. Fermi 5, 50125 Firenze, Italy\\
$^3$Department of Astronomy, Yale University, New Haven, CT 06520-8101, U.S.A.}

\pagerange{\pageref{firstpage}--\pageref{lastpage}}
\pubyear{2001}

\begin{document}

\maketitle

\label{firstpage}

\begin{abstract}
Recent theoretical investigations have suggested that
the formation of the very first stars, 
forming out of metal-free gas, was fundamentally different from
the present-day case.
The question then arises which effect was responsible for
this transition in the star formation properties.
In this paper, we study the
effect of metallicity on the evolution of the gas in a collapsing dark matter 
mini-halo. We model such a system as an
isolated 3$\sigma$ peak of
mass $2\times 10^{6}M_{\odot}$ that collapses at $z_{coll}\simeq 30$,
using smoothed particle hydrodynamics. The gas has a supposed level
of pre-enrichment of either $Z=10^{-4}Z_{\odot}$ or $10^{-3}Z_{\odot}$.
We assume that H$_{2}$ has been radiatively destroyed by the presence
of a soft UV background. Metals therefore provide the only viable
cooling at temperatures below $10^{4}$K.
We find that the evolution proceeds very differently for the two
cases. The gas in the lower metallicity simulation fails to undergo
continued collapse and fragmentation, whereas the gas in the higher
metallicity case dissipatively settles into the centre of the dark
matter halo. The central gas, characterized by densities $n_{\rmn H}
\ga 10^{4}$ cm$^{-3}$, and a temperature, $T\simeq 90$ K, which closely
follows that of the cosmic microwave background,
is gravitationally unstable
and undergoes vigorous fragmentation.
We discuss the physical reason for the existence of a critical metallicity,
$Z_{crit}\sim 5\times 10^{-4}Z_{\odot}$, and its possible dependence
on redshift.
Compared to the pure H/He case, the fragmentation of the $Z=10^{-3}Z_{\odot}$
gas leads to a larger relative number of low-mass clumps.
\end{abstract}


\section{Introduction}
One of the grand challenges in modern astrophysics is to understand how the cosmic
``dark ages'' ended, and to elucidate the nature of the first luminous objects
(e.g., Rees 1999; Barkana \& Loeb 2001). Recently, various authors have addressed the properties
of the very first generation of stars through numerical simulations of the collapse and
fragmentation of primordial clouds, consisting of dark matter (DM) and metal-free, pure
H/He gas (Abel et al. 1998; Bromm, Coppi, \& Larson 1999, 2001; Nakamura \& Umemura 1999;
Abel, Bryan, \& Norman 2000). In the context of hierarchical models of structure formation,
these so-called Population III stars are expected to form in halos of mass
$\sim 10^{5} - 10^{6} M_{\odot}$, collapsing at $z \simeq 20 - 30$ (Tegmark et al. 1997).
The three-dimensional numerical investigations have suggested that Population III
stars were rather massive, perhaps even very massive with $M_{\ast} \geq 100 M_{\odot}$
(Bromm et al. 1999, 2001; Abel et al. 2000).
Based on two-dimensional studies, Nakamura \& Umemura (2001) have hypothesized a bimodal
initial mass function (IMF) for the first stars, with one mode favoring very massive stars,
and the other leading to `normal' stars of mass $\sim 1 M_{\odot}$. The existence of
a high characteristic mass scale has a robust explanation in the microphysics of cooling
due to H$_{2}$ which is the main coolant in the absence of metals.

Although important uncertainties remain, mostly deriving from the poorly understood physics
of the protostellar feedback, it seems plausible that the formation of Population~III
stars was qualitatively different from present-day  (Population~I) star formation.
The question, therefore, arises:
{\it How did the transition in the star formation
properties take place, and what is the physics responsible for it?}
In this paper, we investigate the arguably most important effect:
the presence of a trace amount of heavy elements.
These metals were likely injected into the gas during a pre-enrichment
event, since self-enrichment is
rendered highly implausible due to the blowaway of any remaining gas after the first
(Pop.~III) stars have exploded as SNe, given the relatively small
binding energy of Population III DM halos (Ciardi et al. 2000).

We here consider the limiting case in which all the H$_{2}$
has been radiatively destroyed by the presence of a soft UV background.
H$_{2}$ molecules are very vulnerable to photons with energies below
the Lyman limit. Even before the universe has been reionized, a pervasive,
soft UV radiation field could therefore have been established (e.g., Ciardi et al. 2000).
Metals consequently provide the only cooling 
at temperatures $T\la 10^{4}$K.

Other important roles in modifying the physics of star formation might be
attributable to magnetic fields, or to the onset of turbulence. Both these
effects are believed to be unimportant in the formation of the very first 
stars (e.g., Loeb 1998), but it will be interesting to explore their role
in forming the second generation of stars in future work.

Previous work on this question has focused on the chemical and thermal history of the
gas, employing a simple one-zone model for the dynamical evolution of the collapsing
cloud (e.g., Silk 1977; Yoshii \& Sabano 1980; Lin \& Murray 1992;
Nishi \& Tashiro 2000; Omukai 2000).
Following the collapse 
to the point where the gas becomes opaque, Omukai (2000) argues that the resulting mass of
the protostellar core is independent of metallicity.
The final stellar mass, however, is expected to have no direct
relation to the initial (core) mass, and is mainly determined by the
continued accretion onto the protostellar core.
Our approach, based on full three-dimensional numerical simulations of a collapsing
primordial cloud, allows for the creation of sink particles and is thus
well suited to address the complex physics of accretion.

The paper is organized as follows.
In \S 2, we present our numerical methodology. The choice and justification for the
initial conditions, and the results of the simulation are discussed in \S 3. We summarize
our findings, and address their implications in the final section.

\section{Numerical Method}
The evolution of the dark matter and gas components is calculated with
a version of TREESPH (Hernquist \& Katz 1989), combining the smoothed
particle hydrodynamics (SPH) method with a hierarchical (tree) gravity
solver (see Bromm, Coppi, \& Larson 2001 for further details).
Here, we discuss the additions to the code which are necessary for the
investigation of low metallicity gas.
These are a method to treat the radiative cooling of the gas, 
and a technique to create sink particles.
The thermal evolution of the gas is governed by the equation:

\begin{equation}
\frac{{\rmn D}u}{{\rmn D}t}=\frac{P}{\rho^{2}}\frac{\rmn{D}\rho}{\rmn{D}t}
+ \frac{\Gamma-\Lambda}{\rho}
\end{equation}
 where
 ${\rmn D}/{\rmn D}t$ is the usual Lagrangian time derivative,
 $P$ and $\rho$ are the gas pressure and density, $u$ is the specific
internal energy (in erg g$^{-1}$),
and $\Gamma$ and $\Lambda$ are the contributions from radiative heating and
cooling, respectively (in units of erg cm$^{-3}$ s$^{-1}$).
The first term on the right-hand side describes adiabatic
cooling due to expansion or heating due to compression. We now discuss the
relevant radiative processes.

\begin{figure}
 \vspace{2pt}
 \psfig{file=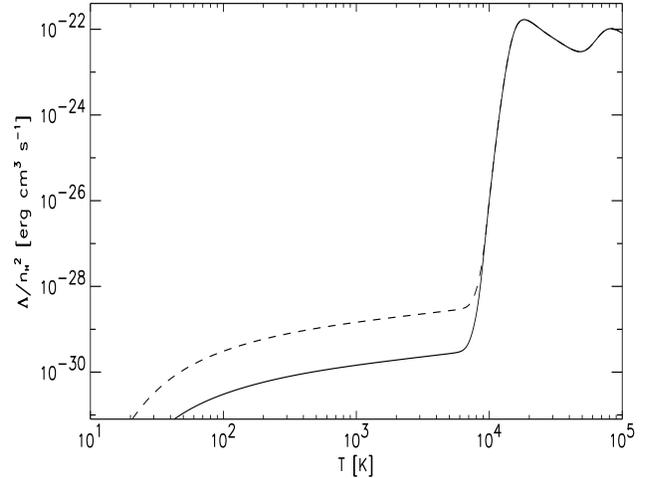,width=8.5cm,height=7.cm}
 \caption{Cooling function for gas which is enriched
with a trace amount of metals. Shown is the cooling rate (in units
of erg cm$^{3}$ s$^{-1}$) vs. temperature, assuming a hydrogen density of $n_{\rmn H}
=1$ cm$^{-3}$.
{\it Solid line:} $Z=10^{-4}Z_{\odot}$.
{\it Dashed line:} $Z=10^{-3}Z_{\odot}$.
  }
\end{figure}

We use the cooling function of Ricotti, Ferrara \& Miniati
(1997) which includes cooling by fine structure and metastable
lines of C, N, O, Fe, S, and Si. Ionization equilibrium
is supposed to be maintained by cosmic rays that we assume to be
associated with the heavy element production by SNe.
The (primary) cosmic ray ionization rate is scaled 
from the Galactic value, $\zeta_{CR} = 1.8 \times 10^{-17}$~s$^{-1}$
(McKee 1995), by the factor $Z/Z_{\odot}$, where $Z_{\odot}$ is the solar
metallicity. The same scaling with $Z$ is used to derive the dust-to-gas ratio,
which we take locally to be equal to 1/160. Again, the rationale is that
dust at high redshift can only be formed in the ejecta of primordial 
Type II SNe (Todini \& Ferrara 2001); hence, 
its production occurs simultaneously with the injection of heavy
elements into the intergalactic medium (IGM).
Dust, however, is only
marginally important for the cooling (via electron recombinations on
positively charged grains) at the low heavy element abundances we
are interested in. In Figure 1, we show the resulting cooling function 
for two different metallicities.

Since radiative cooling to temperatures below that of the cosmic microwave
background (CMB), $T_{\rmn CMB}=2.7\mbox{\,K}(1+z)$, is thermodynamically
not possible, we write for the cooling term
\begin{equation}
\Lambda=\Lambda(T)-\Lambda(T_{\rmn CMB})\mbox{\ \ \ .}
\end{equation}
For $T < T_{\rmn CMB}$, radiative cooling consequently turns into heating.
This approximate treatment ensures that $T\geq T_{\rmn CMB}$, unless cooling proceeds via
adiabatic expansion.
At $z\ga 30$, gas temperatures are therefore limited to $T \ga 90$ K, justifying
the neglect of cooling due to molecules, such as CO, which would become important only
at lower temperatures.

Due to the assumed complete destruction of H$_{2}$
in the presence of a soft UV field, hydrogen molecules do not contribute
to the cooling of the gas.
Prior to reionization there are almost no ionizing photons, and we consequently ignore
heating due to photo-ionization.

We include heating due to the photo-electric effect on dust
grains. This is consistent with our assumption that all the H$_{2}$ has been
destroyed by the presence of a soft UV background, since the photo-electric
heating is mostly due to photons in the same spectral range.
We write the heating due to the photo-electric effect as
$\Gamma=a n$, with $a=a_{\odot}(Z/Z_{\odot})$. For the local
value of the heating constant, we choose $a_{\odot}\simeq
10^{-24}\epsilon$ erg s$^{-1}$, with a heating efficiency of $\epsilon\simeq
0.05$ (Wolfire et al. 1995). As we discuss later, this effect is not very important
for the evolution of the gas.

We have devised an algorithm to merge SPH particles in high-density regions
in order to overcome the otherwise prohibitive time-step limitation, as
enforced by the Courant stability criterion. To follow the simulation
for a few dynamical times, we allow SPH particles to merge into more
massive ones, provided they exceed a predetermined density threshold,
$n_{th}\simeq 10^{6}$ cm$^{-3}$.
More details of the merging algorithm are given in Bromm et al. (2001).

\section{The Simulations}

\subsection{Initial Conditions}

Within a hierarchical cosmogony, the
very first stars are expected to form out
of $3-4 \sigma$ peaks in the random field of primordial density fluctuations.
The early (linear) evolution of such a peak, assumed to be an isolated and
roughly spherical overdensity, can be described by the top-hat
model (e.g., Padmanabhan 1993). We use the top-hat approximation to
specify the initial DM configuration, where we choose
the background universe to be described by the critical Einstein-de Sitter
(EdS) solution with density parameters: $\Omega_{\rmn DM}=0.95$, $\Omega_{\rmn B}=0.05$, and 
Hubble constant $h=H_{0}/(100$ km s$^{-1}$ Mpc$^{-1}$)=0.5.
In this paper, we investigate the fate of a $3 \sigma$ peak
of total mass $2\times 10^{6}M_{\odot}$, corresponding to $10^{5}M_{\odot}$
in baryons which is
close to the
cosmological Jeans mass (Couchman \& Rees 1986).
On this mass scale, the
standard Cold Dark Matter (CDM) scenario predicts a present-day r.m.s. 
overdensity of $\sigma_{0}(M)\simeq 16$, with a normalization
$\sigma_{8}=1$ on the $8h^{-1}$Mpc scale. Then, one can estimate the
redshift of collapse (or virialization) from:
$1+z_{coll}=3 \sigma_{0}(M)/1.69$, leading to $z_{coll}\simeq 30$.

Recent observations (see Barkana \& Loeb 2001 and references therein)
have presented evidence for a flat universe with $\Omega_{m}=1-
\Omega_{\Lambda}=0.3$, thus rendering the EdS universe increasingly
implausible. At high redshifts ($z\ga 10$), however, the standard
CDM model deviates only slightly from the more realistic $\Lambda$CDM
case, and serves as a useful template for all hierarchical models
of structure formation.
The results presented in this paper mainly rely on the
cooling physics of the gaseous component, and are not likely to change
significantly, if these more realistic cosmological parameters
are adopted.

Our simulation is initialized at $z_{i}=100$, by performing the following steps.
The collisionless DM particles are placed on
a cubical Cartesian grid, and are then perturbed according
to a given power spectrum $P(k)=A k^{n}$, by applying the 
Zeldovich approximation (Zeldovich 1970) which 
also allows to self-consistently
assign initial velocities. The power-law index is set to $n=-3$ which
is the asymptotic small-scale 
behaviour
in the standard CDM model (Peebles 1993).
To fix the amplitude $A$, we specify the initial variance of the
fluctuations
\begin{equation}
\sigma_{i}^{2}=A\sum k^{n}
\mbox{\ \ \ .}
\end{equation}
The summation is over all contributing modes, where the minimum
wavenumber is given by the overall size of the Cartesian box, and
$k_{max}$ by the Nyquist frequency. Choosing $\sigma_{i}^{2}\simeq 0.1$,
the rms fluctuation at the moment of collapse is
\begin{equation}
\sigma(z=30)=\left(\frac{1+z_{i}}{1+z}\right)\sigma_{i}\simeq 1
\mbox{\ \ \ .}
\end{equation}
This choice ensures that the substructure develops on a similar
timescale as the overall collapse of the background medium.
Next, particles within a (proper) radius of $R_{i}=$ 150 pc 
are selected for the
simulation. The resulting number of DM particles is here $N_{\rmn DM}=14123$.
Finally, the particles are set into rigid rotation and are endowed
with a uniform Hubble expansion
(see also Katz 1991).
Angular momentum is added by assuming a spin-parameter
$\lambda=L|E|^{1/2}/(G M^{5/2})=0.05$,
where $L$, $E$, and $M$ are the total angular momentum, energy, and mass,
respectively.

The collisional SPH particles ($N_{\rmn SPH}=65536$) are randomly
placed to approximate
a uniform initial density. The random sampling inevitably introduces
shot-noise. 
The DM particles were set up on a regular grid specifically to avoid the
unphysical shot-noise distribution which could mask the desired
physical power spectrum. For the gas, however, the presence
of the shot noise is not a big problem, since the gas mass is initially
slightly smaller than the Jeans mass. Therefore, sound waves will efficiently wipe out
all initial density disturbances.
The SPH particles
are endowed with the same Hubble expansion and rigid rotation as the DM ones.
For the initial gas temperature,
we adopt the 
value (see Tegmark et al. 1997): 
$T_{gas, i}\simeq 200$ K.

We carry out two simulations: Run A having a metallicity of $Z=10^{-3}Z_{\odot}$
, and Run B one of $Z=10^{-4}Z_{\odot}$.
These two values bracket those derived from QSO absorption line studies of the low
column density Ly$\alpha$ forest at $z\sim 3$ (Pettini 1999). They are also close to
those expected to be associated with the reionization process.
Otherwise, the two runs have
identical initial conditions.

\subsection{Results}

The evolution of the collapsing clouds separates into two distinct stages.
Leading up to the virialization of the DM, the two simulations behave
very similarly, and the evolution is rather insensitive to metallicity.
Once the gas has reached a roughly pressure-supported state in the 
potential of the DM halo, the subsequent evolution bifurcates, resulting in
continued collapse and fragmentation for Run A, and in failure to do so
for Run B. We now discuss these two stages in turn.

\subsubsection{Virialization of Dark Matter Halo}
In response to the initially imprinted density fluctuations, the DM develops
a marked substructure. The gas does not `feel' the potential wells of
these subcondensations, however, since it cannot cool sufficiently
during these early evolutionary phase.
Instead, due to adiabatic compression, $T\propto n^{2/3}$, the gas reaches
temperatures of $\sim 10^{4}$ K. At this point, very efficient cooling
due to the excitation of hydrogen lines sets in (see Fig. 1), and maintains
the gas at this temperature upon further compression. At the end of
the virialization process, the gas has reached a state of rough pressure
support with typical gas temperatures close to the virial temperature
\begin{equation}
T_{vir}=\frac{G M m_{\rmn H}}{2 k_{\rmn B} R_{vir}}\sim 5000 \mbox{\ K}
\end{equation}
 where $R_{vir}\simeq 100$ pc is the
 virial radius, $G$ is Newton's constant, $k_{\rmn B}$
Boltzmann's constant, and $m_{\rmn H}$ the mass of a hydrogen atom.
To estimate the corresponding gas density, $n_{vir}$, we consider
the baryonic Jeans mass
\begin{equation}
M_{J}=3\times 10^{7}M_{\odot}\left(\frac{T}{5000 \mbox{\ K}}\right)^{3/2}
\left(\frac{n_{\rmn H}}{1\mbox{\ cm}^{-3}}\right)^{-1/2}
\end{equation}
and assume that it has to be approximately equal to the total mass of the halo.
Here, $n_{\rmn H}$ denotes the hydrogen number density.
From imposing $M_{J}\sim 2\times 10^{6}M_{\odot}$, one then finds
$n_{vir}\simeq 10^{2.5}$ cm$^{-3}$. Heating due to the photo-electric
effect is never important, neither during virialization, nor during
the later, baryon-dominated stages. From here on, the further
evolution strongly depends on the level of trace metal enrichment.

\begin{figure}
 \vspace{2pt}
 \psfig{file=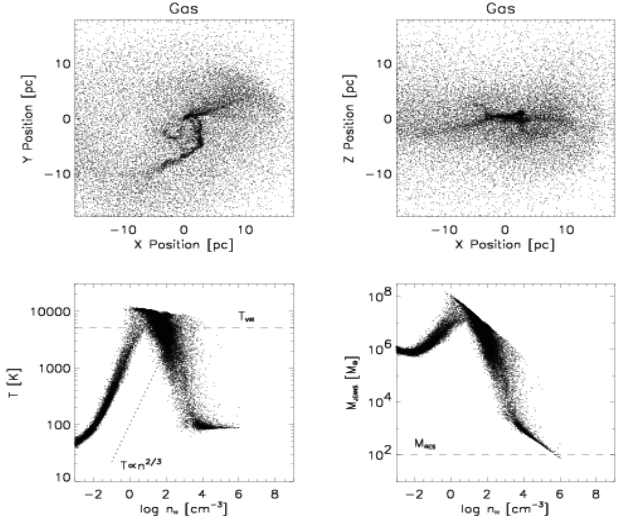,width=8.5cm,height=7.cm}
 \caption{Run A: Morphology
 and gas properties at $z=30.5$. {\it Top row:} The remaining
gas in the diffuse phase.
{\it Left panels:} Face-on view. {\it Right panels:} Edge-on view.
The size of the box is 30 pc. It can be seen that the gas has
dissipatively settled into a central, disk-like configuration.
 {\it Lower-left panel:} Temperature vs. hydrogen number density.
 {\it Lower-right panel:} Jeans mass (in units of $M_{\odot}$) vs. hydrogen number density (in units of cm$^{-3}$).}
\end{figure}

\begin{figure}
 \vspace{2pt}
 \psfig{file=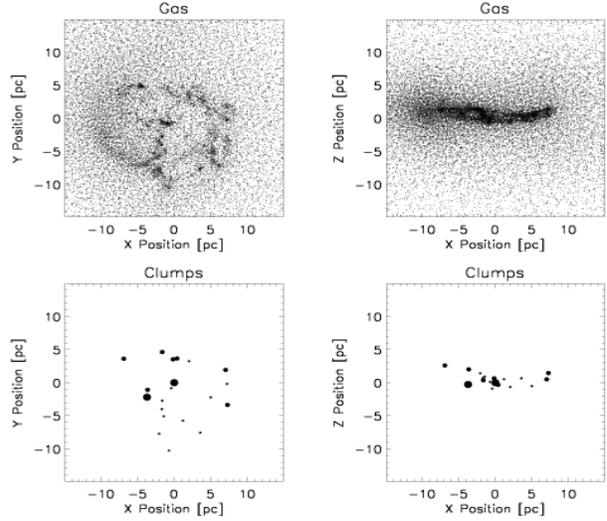,width=8.5cm,height=7.cm}
 \caption{Run A: Morphology at $z=28$. {\it Top row:} The remaining
gas in the diffuse phase.
 {\it Bottom row:} Distribution of clumps. Dot sizes are proportional to the
mass of the clumps.
{\it Left panels:} Face-on view. {\it Right panels:} Edge-on view.
The size of the box is 30 pc. It can be seen that the gas has
undergone vigorous fragmentation.}
\end{figure}

\subsubsection{Further Collapse and Fragmentation of the Gas}
We first discuss Run A. Cooling in this case is efficient enough to
allow further collapse. In Figure 2, we show the gas morphology and
the corresponding thermodynamic state at $z=30.5$. As can be seen,
the gas has dissipatively settled into a disk-like central configuration.
This disk is horizontally supported by rotation. In the presence of a DM
halo, one expects a contraction by a factor of $1/\lambda\simeq 20$ for 
rotational support. The size of the disk, as shown in Figure 2, is in
good agreement with this prediction. 
It is also evident that the disk is subject to a bar-mode ($m=2$)
instability. By examining the $T-n_{\rmn H}$ plane (lower-left panel in Fig. 2), one
can see that the gas efficiently cools from $\sim 10^{4}$ K down to
the value set by the CMB floor, $T_{min}\simeq T_{\rmn CMB}=86$ K. During this
rapid cooling, the gas remains rather smooth. The gravitationally
unstable disk-material, however, subsequently
undergoes vigorous fragmentation (see Figure 3).

We next turn to Run B, and to its rather different fate. In Figure 4, we show
the situation at $z\simeq 28$. In contrast to the corresponding situation
for Run A (Figure 3), no further collapse, and no fragmentation has
occurred. Instead, the gas remains in pressure support, and lingers at
temperatures close to $T_{vir}$. Only much later, as the result of
slow but inexorable residual cooling, two high density clumps form at the
very centre of the roughly spherical gas configuration (see Figure 5).
Run B, therefore, presents an example of what Ciardi et al. (2000) have
termed `dark object', a halo which has failed to produce anything luminous
in it. 

To summarize, in Run A, the gas becomes effectively self-gravitating,
leading to continued collapse and fragmentation, whereas in Run B, the
gas does not. To see this more clearly, we compare in Figure 6 the cooling
timescale, $t_{cool}\simeq(n_{\rmn H} k_{\rmn B} T)/\Lambda$, with the freefall
time, $t_{ff}\simeq \left(G m_{\rmn H} n_{\rmn H}\right)^{-1/2}$ (Rees \& Ostriker 1977).
Whether the gas can reach higher densities due to its self-gravity is
decided at the end of the DM virialization process, with the gas having
$T\sim 5000$ K and $n_{\rmn H}\sim 10^{2.5}$ cm$^{-3}$. For these values, one
has $t_{cool}\simeq t_{ff}$ for Run A, and $t_{cool} > t_{ff}$ for
Run B. Consequently, the gas in Run B is not able to cool any further and to reach
higher densities.

We now give an analytical estimate for the value of the
critical metallicity, $Z_{crit}$, which discriminates between the two
cases discussed above.
Evaluating the cooling function (see Fig. 1) at $T=5000$ K, we have
\begin{equation}
\Lambda (Z)\simeq 3\times 10^{-30}\mbox{erg cm$^{-3}$ s$^{-1}$}
\left(\frac{Z}{10^{-4}Z_{\odot}}\right)
\left(\frac{n_{\rmn H}}{1 {\rmn cm}^{-3}}\right)^{2}{\rmn \ .}
\end{equation}
The adiabatic heating,
$\Gamma_{ad}=\frac{P}{\rho}\frac{\rmn{D}\rho}{\rmn{D}t}\simeq \frac{P}{t_{ff}}$,
can be written as
\begin{equation}
\Gamma_{ad}\simeq 10^{-24}\mbox{erg cm$^{-3}$ s$^{-1}$}
\left(\frac{T}{5000 {\rmn K}}\right)
\left(\frac{n_{\rmn H}}{300 {\rmn cm}^{-3}}\right)^{3/2}{\rmn \ .}
\end{equation}
Equating the two terms yields $Z_{crit}\simeq 5\times 10^{-4}Z_{\odot}$,
in accordance with the results of our simulations.

We finally address the dependence of $Z_{crit}$ on the redshift of
collapse, $z_{coll}$, and the total mass of the halo, $M$.
The gas temperature and density at the end of the DM virialization
are expected to obey the relations
$T\sim T_{vir}\propto M^{2/3}(1+z_{coll})$ and
$n_{\rmn H} \propto \rho_{vir}\propto (1+z_{coll})^{3}$, respectively
(e.g., Padmanabhan 1993). Assuming $\Lambda \propto T^{1/2}$ for
$10^{2} \la T \la 10^{4}$ K, 
inserting the virial relations into equations
(7) and (8), and normalizing to the case treated in this paper, one
approximately finds
\begin{equation}
Z_{crit} \simeq 5\times 10^{-4}Z_{\odot}
\left(\frac{M}{2\times 10^{6}M_{\odot}}\right)^{\frac{1}{3}}
\left(\frac{1+z_{coll}}{31}\right)^{-1}{\rmn \ .}
\end{equation}
This relation is only valid for halos with $T_{vir} < 10^{4}$ K.
We plan to test this analytical prediction with future simulations.

\begin{figure}
 \vspace{2pt}
 \psfig{file=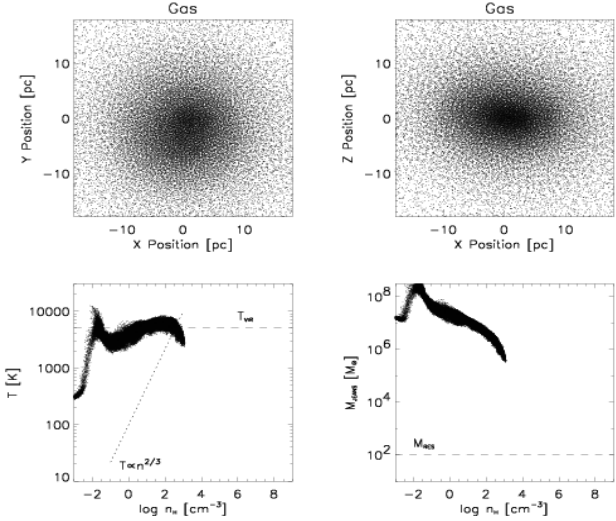,width=8.5cm,height=7.cm}
 \caption{Run B: Morphology and thermodynamic properties of the gas
at $z=28$.
The convention of Fig. 2 is adopted for the rows and columns.
}
\end{figure}

\begin{figure}
 \vspace{2pt}
 \psfig{file=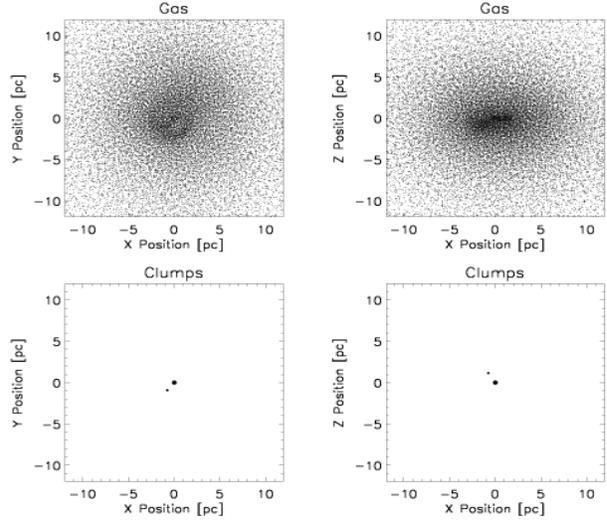,width=8.5cm,height=7.cm}
 \caption{Run B: Morphology 
at $z=22.7$. The convention of Fig. 3 is adopted for the rows and columns.
}
\end{figure}

\begin{figure}
 \vspace{2pt}
 \psfig{file=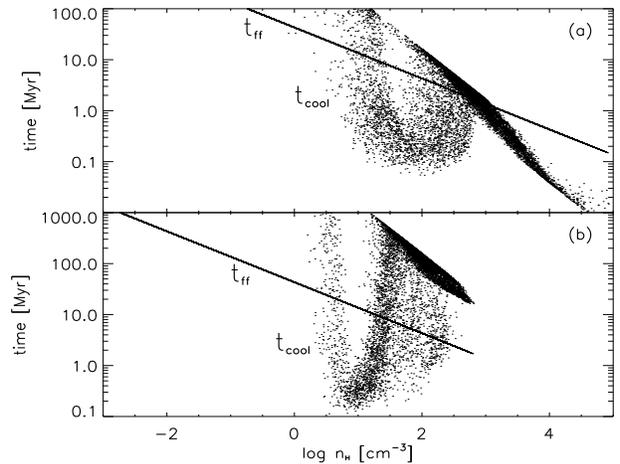,width=8.5cm,height=7.cm}
 \caption{The criterion for continued collapse. Freefall timescale
({\it solid line}) and cooling timescale ({\it dotted symbols})
vs. hydrogen number density (in cm$^{-3}$). Timescales are in units of $10^{6}$ yr.
Shown is the situation briefly after the virialization of the DM, at
$z\simeq 30.5$.
{\bf (a)} Run A: The cloud succeeds in undergoing continued collapse due
to efficient cooling, $t_{cool} \sim t_{ff}$, at log $n_{\rmn H}\simeq 2.5$.
{\bf (b)} Run B: Here, $t_{cool} > t_{ff}$
at log $n_{\rmn H}\simeq 2.5$, and the gas fails to undergo continued
collapse.
}
\end{figure}

\subsubsection{Comparison with the $Z=0$ Case}
Since Bromm et al. (1999,2001) have investigated the case of zero-metallicity
gas with initial conditions that are otherwise
very similar to those adopted here, it
is instructive to compare our $Z=10^{-3}Z_{\odot}$ simulation with the
pure H/He case.
In the previous papers, the formation of H$_{2}$ was allowed for, whereas here
we have considered the limiting case of maximum negative feedback which is supposed 
to destroy all H$_{2}$.

The first difference occurs during the virialization of the DM. In marked
contrast to the simulations presented in this paper, the DM fluctuations
do imprint their signature onto the gaseous component if H$_{2}$ is the
main coolant. These DM induced condensations act as the seeds for the
subsequent fragmentation of the gas.
The basic reason for this difference is the boost in the formation of H$_{2}$
due to the density and temperature enhancement in the DM subcondensations.
Such a chemistry-related feedback is absent in the $Z=10^{-3}Z_{\odot}$ case.

The second, crucial, difference derives from the fact that cooling
remains efficient at densities $n_{\rmn H}\ga 10^{4}$ cm$^{-3}$, provided
the gas has been enriched to $Z\geq Z_{crit}$. As can be seen in
Figure 2 (lower-left panel), the gas evolves isothermally up
to the threshold density of $n_{th}=10^{6}$ cm$^{-3}$. H$_{2}$ cooling, on
the other hand, saturates at $n_{\rmn H}\ga n_{crit}\simeq 10^{3}-
10^{4}$ cm$^{-3}$, where $n_{crit}$ marks the transition from NLTE
to LTE level populations. By comparing the distribution of
$M_{J}$ vs. $n_{\rmn H}$ in Figure 2 (lower-right panel) to the
equivalent Figure 3 (panel (c)) in Bromm et al. (1999), it can be seen
that there is no characteristic mass scale in
the $Z=10^{-3}Z_{\odot}$ case.
Such a characteristic scale would correspond
to a pile-up of SPH particles in the diagram
due to long evolutionary timescales.
In the pure H/He case, on the other hand, there is
a preferred mass scale of $M \sim 10^{3}M_{\odot}$.

The overall range of clump masses, $10^{2}<M_{Cl}<10^{4}M_{\odot}$,
is similar in the pure H/He and $Z=10^{-3}Z_{\odot}$ simulations, although
the relative number of low-mass clumps
is significantly higher for the pre-enriched case. These low-mass clumps have
masses close to the resolution limit of the simulation
(Bate \& Burkert 1997)
\begin{equation}
M_{res}\simeq M_{\rmn B}
\left(\frac{2 N_{neigh}}{N_{\rmn SPH}}\right)\simeq 100 M_{\odot}\mbox{\ ,}
\end{equation}
where $N_{neigh}\simeq 32$ denotes the number of SPH particles within a
smoothing kernel. Obviously, clumps cannot have masses smaller than $M_{res}$.
The most massive clumps form with initial masses close to the Jeans mass,
$M_{J}\sim 750 M_{\odot}$, evaluated at the density, $n_{\rmn H}\simeq
10^{4}$ cm$^{-3}$, where the bulk of the freely-falling gas reaches the CMB
temperature floor, $T_{min}\simeq 86$ K (see Fig. 2).

Later on, clumps gain in mass through accretion from surrounding gas,
and through occasional mergers with other clumps. 
An investigation of the possible subfragmentation of a clump lies
beyond the resolution limit of our simulation.
It is worth emphasising that the spectrum of clumps is different
from the {\it stellar} spectrum (i.e., the IMF), and it is not possible to
directly derive the latter from the former. However, the qualitative trend
that the relative number of low-mass clumps increases with
metallicity is suggestive of
the possible regulation of the IMF by the gas metal enrichment.
To further constrain the clump mass spectrum, and eventually the stellar IMF,
higher resolution studies will be necessary.

The fact that the
minimum temperature enforced by the CMB could imply larger
Jeans masses at high redshift has already been pointed out by Larson (1998).
The role of the CMB in setting the minimum temperature points to the
possible importance of $z_{coll}$ in setting an evolving mass scale
for cosmic star formation (see Bromm \& Clarke 2001 for a discussion
of this effect).


\section{Summary and Conclusions}
We have discussed the evolution of primordial star-forming objects in the
context of the standard CDM model for structure formation. The studied clouds are
pre-enriched to a level of $10^{-4}$ and $10^{-3}Z_{\odot}$, consistent with
observations of the IGM, and collapse close to $z_{coll}\sim 30$.
We find that the evolution proceeds very differently in the two cases.
The gas in the higher metallicity run can efficiently cool, and consequently
collapse to a disk-like configuration. The disk material is gravitationally
unstable and fragments into a large number ($\sim$ 25) of high-density
clumps. In marked contrast to this case, the gas in the lower metallicity
simulation cannot efficiently cool, and therefore remains in the
post-virialization, pressure-supported state in a roughly spherical
configuration. Such a system constitutes an example of a `dark object', as discussed
by Ciardi et al. (2000).
These results indicate the existence of a critical
metallicity, $Z_{crit}\simeq 5\times 10^{-4}Z_{\odot}$, below which
the presence of heavy elements does not significantly influence the
evolution of a primordial cloud. The value of the critical metallicity might
be parameter dependent, being larger at lower $z$ and in more massive halos.
To account for the observed abundance pattern in metal-poor stars,
Wasserburg \& Qian (2000) have hypothesized an initial, prompt enrichment
episode due to a generation of very massive stars. According to their
model, these stars would have formed in gas with [Fe/H]$\la -3$. This
nucleosynthetic evidence for a critical metallicity is consistent with
our result which is based on the physics of star formation.

Our investigation has treated the case of a low mass halo, of mass
$\sim 10^{6}M_{\odot}$, and the question naturally arises of what happens
in more massive ones. Is there still a critical
metallicity, and if so, what would be its numerical value? In addressing
these questions, one can argue that halos with virial temperatures
in excess of $\sim 10^{4}$ K are much less dependent on a level of
pre-enrichment in order to form stars. The gas in these halos
can initially cool through line-excitation of atomic hydrogen, and in the
later stages of molecular hydrogen. Even if not present initially,
or if completely destroyed in the shocks associated with the virialization process,
H$_{2}$ is
expected to be efficiently formed in the post-shock regions of these
systems (e.g., Shapiro \& Kang 1987). Due to their larger binding energy,  
these halos can experience multiple episodes of self-enrichment. Whether
there is a strong change in the star formation properties,
once this internal enrichment exceeds a certain value,
cannot be answered without simulating the evolution of
such a high-$T_{vir}$ halo. 

Although there remain important uncertainties with regard to our
findings, we conclude by discussing their potential implications.
If indeed stars which formed from extremely
low metallicity gas, with $Z < Z_{crit}$, were
predominantly massive, this would entail short stellar-evolutionary timescales,
$t_{evol}\sim 10^{6}$ yr. Consequently, it is not expected to find
any extremely low-metallicity
stars which are still `alive' today. This prediction is consistent with observations
of metal-poor halo stars, having $Z\geq 10^{-4}Z_{\odot}$ (Beers 2000).
Assuming that massive Pop.~III stars end their lives in massive
black holes (MBHs), with typical mass $\sim 10^{3}M_{\odot}$, Madau \& Rees (2001)
have identified high-$\sigma$ peaks of mass $\sim 10^{6}M_{\odot}$, collapsing
at $z\simeq 20-30$, as the sites of their formation. As is suggested in this
paper, MBHs could only have formed early in the history of the universe, out
of gas with $Z < Z_{crit}$.

If stars more massive than $\sim 250
M_{\odot}$ buried their metal production in collapsing to a black
hole (Fryer, Woosley, \& Heger 2001), how did the IGM get enriched
to a level of $Z > Z_{crit}$, enabling the subsequent formation of `normal', low mass
stars? The answer to this question has to await the precise determination of
whether the majority of Population~III stars has masses below or above the threshold
mass for the complete collapse into a black hole, and a better understanding
of how the metals are expelled from the star forming halo (e.g., Madau,
Ferrara, \& Rees 2001).


\section*{Acknowledgments}

We would like to thank Lars Hernquist for making available to us a version
of TREESPH. Support from the NASA ATP grant NAG5-7074 is gratefully
acknowledged. VB thanks the Osservatorio di Arcetri for its hospitality during
the completion of this work. This work has been partially supported by the
EC RTN network ``The Physics of the Intergalactic Medium''.

\end{document}